\documentclass[12pt]{article}

\pdfoutput=1

\usepackage{a4wide}
\usepackage{amssymb}
\usepackage{graphicx}
\usepackage{verbatim}

\usepackage[intlimits]{amsmath}

\usepackage{amsfonts}

\usepackage{epsfig}

\newcommand{\be}{\begin{equation}}
\newcommand{\ee}{\end{equation}}
\newcommand{\bea}{\begin{eqnarray}}
\newcommand{\eea}{\end{eqnarray}}
\newcommand{\ba}{\begin{eqnarray}}
\newcommand{\ea}{\end{eqnarray}}

\def\Xint#1{\mathchoice
   {\XXint\displaystyle\textstyle{#1}}%
   {\XXint\textstyle\scriptstyle{#1}}%
   {\XXint\scriptstyle\scriptscriptstyle{#1}}%
   {\XXint\scriptscriptstyle\scriptscriptstyle{#1}}%
   \!\int}
\def\XXint#1#2#3{{\setbox0=\hbox{$#1{#2#3}{\int}$}
     \vcenter{\hbox{$#2#3$}}\kern-.52\wd0}}
\def\dashint{\Xint-}

\begin{document}

\setcounter{table}{0}

\begin{flushright}\footnotesize

\texttt{ICCUB-20-012}

\end{flushright}

\mbox{}
\vspace{0truecm}
\linespread{1.1}

\vspace{0.5truecm}

\centerline{\Large \bf Deformed Cauchy random matrix ensembles} 
\medskip
\centerline{\Large \bf and large $N$ phase transitions}

\vspace{1.3truecm}

\centerline{
    {\large \bf Jorge G. Russo} }

\vspace{0.8cm}

\noindent  
\centerline {\it Instituci\'o Catalana de Recerca i Estudis Avan\c{c}ats (ICREA), }
\centerline{\it Pg. Lluis Companys, 23, 08010 Barcelona, Spain.}

\medskip
\noindent 
\centerline{\it  Departament de F\' \i sica Cu\' antica i Astrof\'\i sica and Institut de Ci\`encies del Cosmos,}
\centerline{\it Universitat de Barcelona, Mart\'i Franqu\`es, 1, 08028 Barcelona, Spain. }

\medskip

\centerline{  {\it E-Mail:}  {\texttt jorge.russo@icrea.cat} }

\vspace{1.2cm}

\centerline{\bf ABSTRACT}
\medskip

We study a new hermitian one-matrix model
containing a logarithmic Penner's type term and
another term, which can be obtained as a limit from logarithmic terms. For small coupling, the potential has an absolute minimum
at the origin, but beyond  a certain value of the coupling the potential 
develops a double well. For a higher critical value
of the coupling, the system undergoes a large $N$ third-order phase transition.

\noindent

\vskip 1.2cm
\newpage

\def\sech{ {\rm sech}}
\def\p{\partial}
\def\pa{\partial}
\def\ov{\over }
\def\a{\alpha }
\def\g{\gamma}
\def\s{\sigma }
\def\td{\tilde }
\def\vp{\varphi}
\def\strokedint{\int}
\def \ha {{1 \over 2}}

\def\KK{{\cal K}}




\textwidth = 460pt
\hoffset=0pt



\section{Introduction}

The statistical ensembles of random matrices have a vast number of applications
in various domains \cite{mehta,forrester,akemann,baik}.
Originally introduced by Wigner \cite{wigner} -- motivated by the spectral properties of nuclear resonance levels --
they rose in the late 70's as a tool to elucidate aspects of the non-perturbative
structure of QCD in the large $N$ expansion \cite{Brezin:1977sv}. Random matrices can be used to generate sums over
random surfaces, a feature that led to non-perturbative formulations  of two-dimensional gravity
and non-critical strings \cite{DiFrancesco:1993cyw}. More generally, matrix models are very efficient in computations
of combinatorial graph generating functions \cite{Bessis:1980ss,Eynard:2008we}. This property has been exploited in different contexts, including  classifying  numbers of RNA complexes of an arbitrary topology \cite{Bhadola:2013bla,Andersen:2013tsa}.
The topological expansion of the free energy also provides an ideal framework to understand detailed aspects of
resurgent analysis and large $N$ instantons \cite{Pasquetti:2009jg,Aniceto:2011nu,Marino:2012zq}. Large $N$ expansions can be recursively
generated by loop equations \cite{Migdal:1984gj}, which are encoded in a spectral curve.

Large $N$ random matrices, in particular, multi-matrix ensembles,  have also been used to model interesting statistical systems, 
including Ising \cite{Kazakov:1986hu,Boulatov:1986sb} and Potts \cite{Kazakovp,ZinnJustin:1999jg} models,  $O(N)$ models \cite{Kostov:1992pn,Eynard:1992cn}, among others. 
Other important aspects of random matrices include the description in terms of a conformal field theory   and  relations with  integrable hierarchies
(see {\it e.g.} \cite{Kostov:1999xi}).

The Penner model \cite{Penner:1988cza} is a hermitian one-matrix model with a potential containing polynomial terms and logarithmic terms. It has been exhaustively investigated, including the classification  of  critical points to all genera  \cite{Ambjorn:1994bp}.
In this paper we will explore random matrix models arising as a limit of models with potentials having logarithmic singularities in the complex plane.
We will focus on a particular theory of a general class of models having large $N$ quantum phase transition of the third order at a critical
coupling. Large $N$ phase transitions are familiar in random matrix models \cite{Gross:1980he,Wadia:2012fr,Wadia:1980cp}
and singularities typically reflect  the
finite convergence radius of the planar expansion.
The model is introduced in section 2, where the large $N$ limit is studied.
The construction hints that the present matrix model might be exactly solvable at finite $N$, though solving the model is  beyond the scope of this paper.
A discussion of finite $N$ partition functions is included in section 3.


\section{The model }

We  consider an hermitian $N\times N$ matrix $M$
with the dynamics governed by  the potential
\be
 V(M) =A \, {\rm Tr} \ln(1+M^2) + B\, {\rm Tr} \frac{1}{1+M^2}\ .
\label{pot}
\ee
For stability, we must require $A>0$. However, convergence of the partition function will imply the stronger condition $A >N-1$.

\subsection{The 3-parameter deformed Cauchy  model}

Before studying the dynamics of the matrix theory \eqref{pot} for its own sake, it is instructive to elucidate on the
close relation to the familiar Penner random matrix models, where 
the  potential is of the form
\be
 U(x) =U_0(x) -  \sum_{i=1}^n A_i \ln ( x-\alpha_i ) \ ,
\label{penners}
\ee
where $U_0$ is a polynomial (for example, the  Gaussian potential $U_0=c\, x^2$).
In a number of cases, the Penner model can be exactly solvable, with the partition function computed for any $N$, using
Selberg's integral formula \cite{selberg} and generalizations \cite{mehta}. Alternatively, the partition function can be computed by recursion relations using the method of orthogonal polynomials \cite{DiFrancesco:1993cyw,Marino:2004eq} (see \cite{Alvarez:2014fba} for applications of this method to Penner  models).

We consider the potential 
\be
U(x) =  A \ln ( x^2+ 1)+  \beta \ln ( x^2+ 1+\epsilon)-  \beta \ln ( x^2+ 1) \ .
\label{potek}
\ee
This is in the class of Penner models \eqref{penners}  with the choice 
\bea
&& U_0(x)=0 , \quad n=6\ ,\quad  
\alpha_{1,2}=\alpha_{5,6}=\pm i,\quad  \alpha_{3,4}=\pm i \sqrt{1+\epsilon} \ ,\nonumber\\ 
&& A_1=A_2\equiv -A\ ,\quad A_3=A_4=-\beta\ ,\quad A_5=A_6=\beta\ . \nonumber
\eea
The particular case with $A=N$ was recently studied in section 3 of \cite{Santilli:2020ueh}, where 
it originated from a family of unitary matrix  models through the map to the unit circle.
This case is special, as we shall discuss, and in the large $N$ theory corresponds to the marginal case for stability of the model. 
When $\epsilon=0$, the potential \eqref{potek} describes a Cauchy ensemble, studied in \cite{witte,Santilli:2020ueh}.

The potential \eqref{potek} has one absolute minimum at 
$x=0$ for $\epsilon\beta \leq A(1+\epsilon)$ and two minima at
$x=\pm \sqrt{\frac{\epsilon\beta}{A}-1-\epsilon}$ for 
$\epsilon\beta >A(1+\epsilon)$. We will refer to the
 hermitian one-matrix theory with potential \eqref{potek} as the {\it  3-parameter deformed Cauchy model}, 
 or 3-parameter biphasic Cauchy model, since the
 theory has a large $N$ phase transition on a critical line in parameter space. In this paper we will not study the
 phase transitions of this model, but instead focus
 on the matrix theory with potential \eqref{pot}.
 We will return to the 3-parameter biphasic Cauchy model
 in section 3.
 
 \medskip
 
To connect with theory \eqref{pot}, we take the limit $\beta\to\infty$, $\epsilon\to 0$, with fixed $\beta \epsilon \equiv B$. This exactly gives our model \eqref{pot}.
The close relation with Penner models suggests that the model \eqref{pot} could be exactly solvable at finite $N$ (see
 section 3).

\subsection{Partition function at large $N$}

The partition function is given by
\begin{equation}
Z=\frac{1}{{\rm vol}\Big(U(N)\Big)} \int DM \ \exp \left[- V(M) \right]\ .
\end{equation}
After gauge fixing to a diagonal $U(N)$ matrix 
$M={\rm diag}(a_1,...,a_N)$, the partition function becomes
\begin{equation}
Z=\frac{1}{N!} \int \frac{d^Na}{(2\pi)^N} \ \prod_{i<j} (a_i-a_j)^2 \exp \left[-  W(a_i) \right]\ ,
\label{partio}
\end{equation}
where
\be
W(a_i) = \sum_{i=1}^N \left(A \ln (1+a_i^2) + B \frac{1}{1+a_i^2}\right)\ .
\ee
We note that the integral is convergent provided $A> N-1$.
In the large $N$ theory, this condition will naturally arise in order to avoid  an instability in the distribution of eigenvalues.

\begin{figure}[h!]
\centering
\begin{tabular}{cc}
\includegraphics[width=0.4\textwidth]{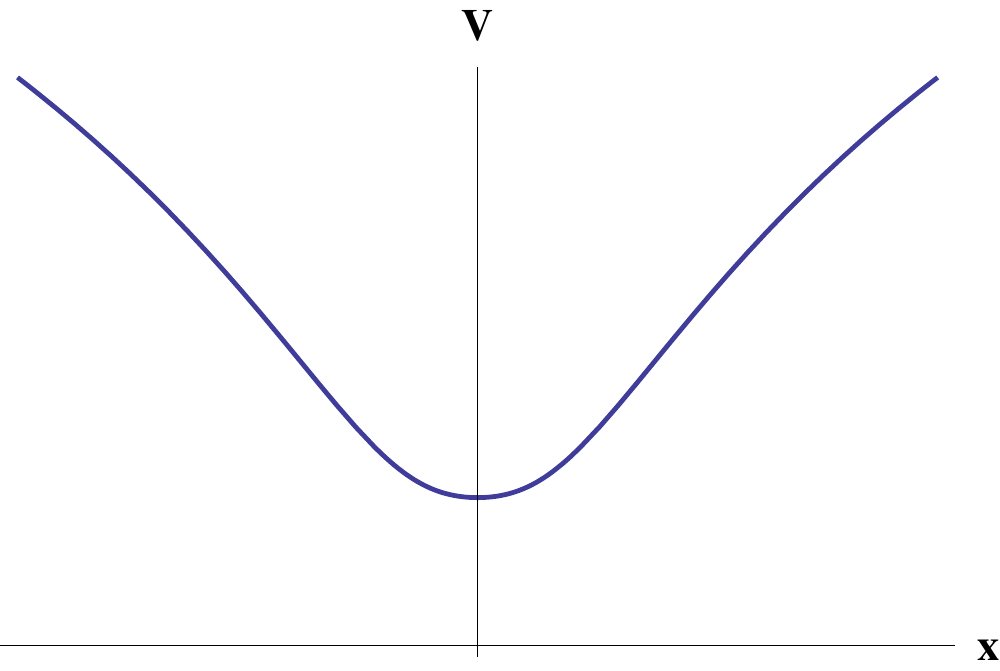}
&
\qquad \includegraphics[width=0.4\textwidth]{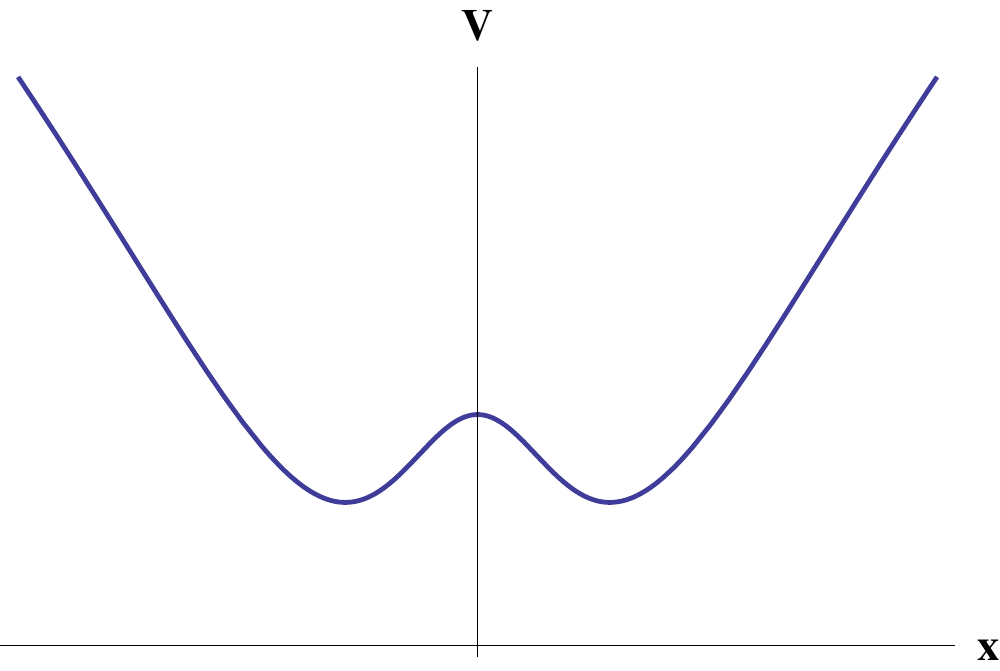}
\\
(a)&(b)
\end{tabular}
\caption
 {The potential has a minimum at $x=0$ for $B<A$  and two  minima at $x=\pm \sqrt{\frac{B}{A} -1}$ for $B>A$. In figure a), $A=2$, $B=1.5$. Figure b) $A=2$, $B=3.5$.}
\label{poteab}
\end{figure}

The dynamics of the system can be intuitively understood as follows.
There is the usual repulsive force between eigenvalues produced
by the Vandermonde determinant. In addition, each eigenvalue is subject to the potential
\be
V= A \ln (1+x^2) + B \frac{1}{1+x^2}\ .
\label{poti}
\ee
The first term, with $A>0$, produces an attractive force that pushes the eigenvalue towards the origin. The second term produces an attractive force for $B<0$ and a repulsive force for $B>0$.
The potential has an absolute minimum at the origin for $B<A$.
However, when $B>A$, in the vicinity of the origin the repulsive force overcomes the attractive force and the vacuum at $x=0$ becomes unstable:  
the potential develops a double well,
with minima at $x=\pm \sqrt{\frac{B}{A} -1}$ (see fig. \ref{poteab}).\footnote{One could introduce a small symmetry breaking term to look for broken
symmetry solutions. We will not attempt this here
(a discussion can be found in \cite{Brower:1992mn}).}

\medskip

We will study the large $N$ limit with $A, B\to\infty$
and fixed couplings
$\tau, \ \kappa$ defined by
\be
\tau \equiv \frac{A}{N}={\rm fixed} \ ,\qquad \kappa\equiv \frac{B}{N}={\rm fixed}\ .
\ee
We introduce the eigenvalue density 
\be
\rho(x) =\frac{1}{N}\sum_{i=1}^N \delta(x-a_i)\ .
\ee
It satisfies  the normalization 
condition
\be
\int_{L} dx\, \rho(x) =1\ ,
\label{normaa}
\ee
where $L$ is the region of the complex $x$ plane where
eigenvalues condense.

In the large $N$ limit, the partition function can be computed exactly by the saddle-point method.
In this limit, the saddle-point equations reduce to the following
singular integral equation
\be
\dashint_{L} dz \frac{\rho(z)}{x-z}= \frac12 \, V'(x)=  \tau\ \frac{x}{1+x^2} -\kappa\ \frac{x}{(1+x^2)^2}\ .
\label{integeq}
\ee

Starting with small $\kappa $, one expects a one-cut eigenvalue distribution, as the potential has an absolute minimum at $x=0$ and
eigenvalues will be pushed towards the minimum.
However, the potential develops a double well 
when $\kappa >\tau $. As $\kappa $ is further increased there should be critical point
where the eigenvalues get separated into a symmetric two-cut distribution. 
This would imply that the system undergoes a phase transition at a critical $\kappa_c$.
In what follows we confirm this picture by explicitly solving the integral equation in the two phases.

\subsection{The one-cut solution}

For sufficiently small $\kappa $, the eigenvalues are expected to condense
in one cut, $-\mu<x<\mu$, with a density satisfying the normalization condition \eqref{normaa} with $L=(-\mu,\mu)$, {\it i.e.} 
\be
\int_{-\mu}^\mu dx\, \rho(x) =1\ .
\label{noma}
\ee
In order to solve the saddle-point equation, as usual one introduces the resolvent
\be
\omega(z) =\int dx\, \frac{\rho(x)}{z-x} \ .
\ee
Then the density is determined by
\be
\rho(x) =-\frac{1}{2\pi i} \left( \omega (x+i\epsilon) -\omega (x-i\epsilon)\right)\ .
\ee
This leads to 
\be
\rho(x) = -\frac{1}{2\pi^2 } \sqrt{\mu^2-x^2}  \int_{-\mu}^\mu dz
\frac{V'(z)}{\sqrt{\mu^2-z^2} (x-z)}\ .
\ee
The integral can be computed by residues by considering a contour
surrounding the cut.
There is no pole at infinity, and the result is given
by the residues at $z=\pm i$.
We obtain the following expression for  the eigenvalue density 
\be
 \rho(x) =\frac{1}{2\pi  (1+\mu^2)^{\frac32}}  \frac{\sqrt{\mu^2-x^2}}{(1+x^2)^2}\, \left(2 \tau (1+\mu^2) (1+x^2) -\kappa (2+\mu^2(1-x^2))\right)\,  \,\ .
\label{rhotot}
\ee

The parameter $\mu$ representing the width of the eigenvalue
distribution is determined by the normalization condition \eqref{noma}. Computing this integral, we obtain the condition
\be
1=  \tau \left(1- \frac{1}{\sqrt{1+\mu^2}}\right)- \frac{\kappa }{2}\,  \frac{\mu^2}{(1+\mu^2)^{\frac32}}\ .
\label{norma}
\ee
This leads to a cubic equation
for $X\equiv \mu^2$.
A valid solution for a  one-cut distribution requires that $\mu^2$ is real and that $\rho(x)$ is non-negative in the interval $-\mu <x< \mu$.
For $\tau  >1$,  the normalization condition always has a real solution for $\mu$, irrespective of the value of $\kappa $.
If $\tau <1$, then
the system is unstable at large $N$; the repulsive (Vandermonde) force of eigenvalues dominate over the attractive force  and eigenvalues spread out to infinity. 
The mathematical origin of the condition $\tau >1$ is the convergence condition 
of the (large $N$) matrix integral \eqref{partio}.\footnote{We thank K. Zarembo for this remark.}
Finally, the case $\tau=1$ is marginal and will be
discussed separately  in section 2.6.

\begin{figure}[h!]
\centering
\begin{tabular}{cc}
\includegraphics[width=0.4\textwidth]{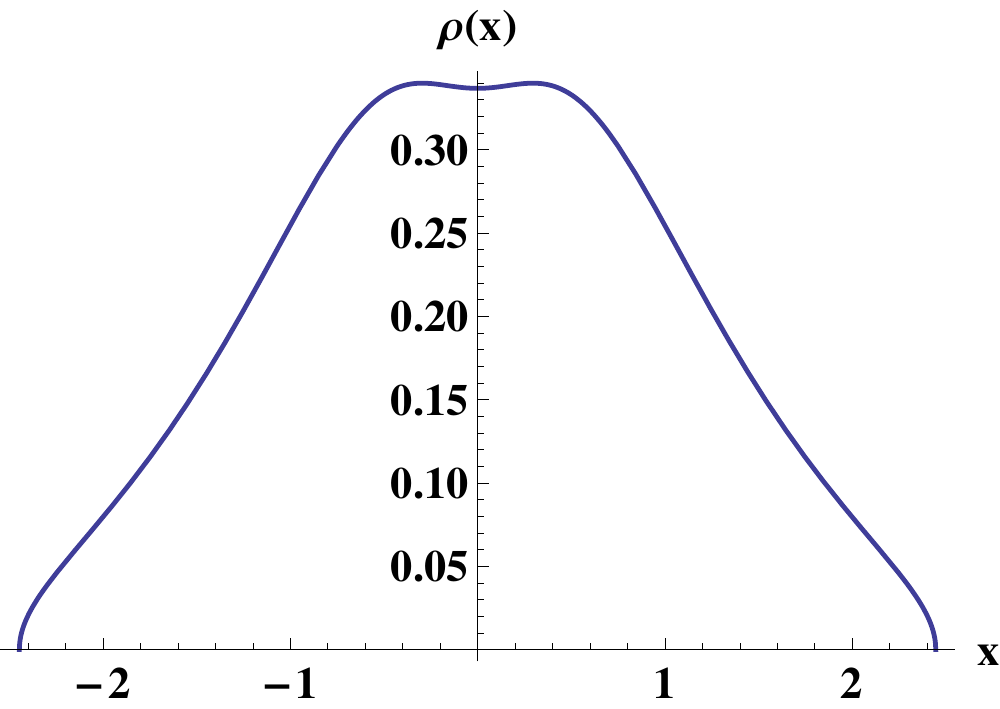}
&
\qquad \includegraphics[width=0.4\textwidth]{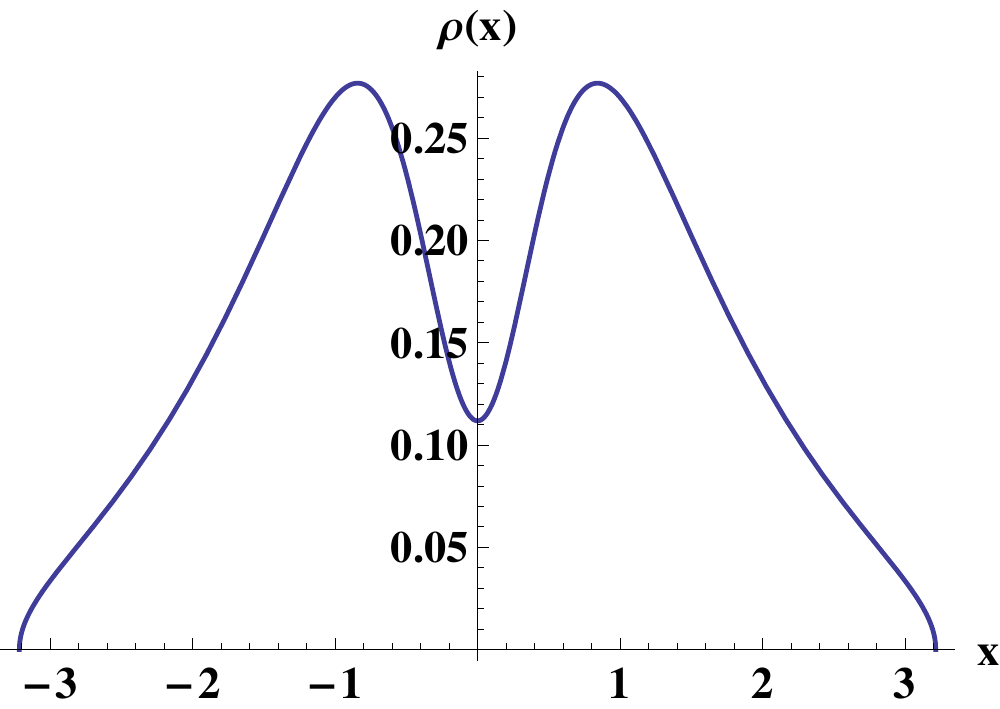}
\\
(a)&(b)
\end{tabular}
\caption
 {The eigenvalue density in the one-cut case describing the subcritical regime $\kappa < \kappa_c$. (a) $\tau=2$, $\kappa =1.5$. (b) $\tau =2$, $\kappa =3$. In this case, the potential has already developed a double-well, but there is an overfilling of eigenvalues, which are still distributed in one cut. }
\label{rhosubc}
\end{figure}

\begin{figure}[h!]
\centering
\includegraphics[width=0.45\textwidth]{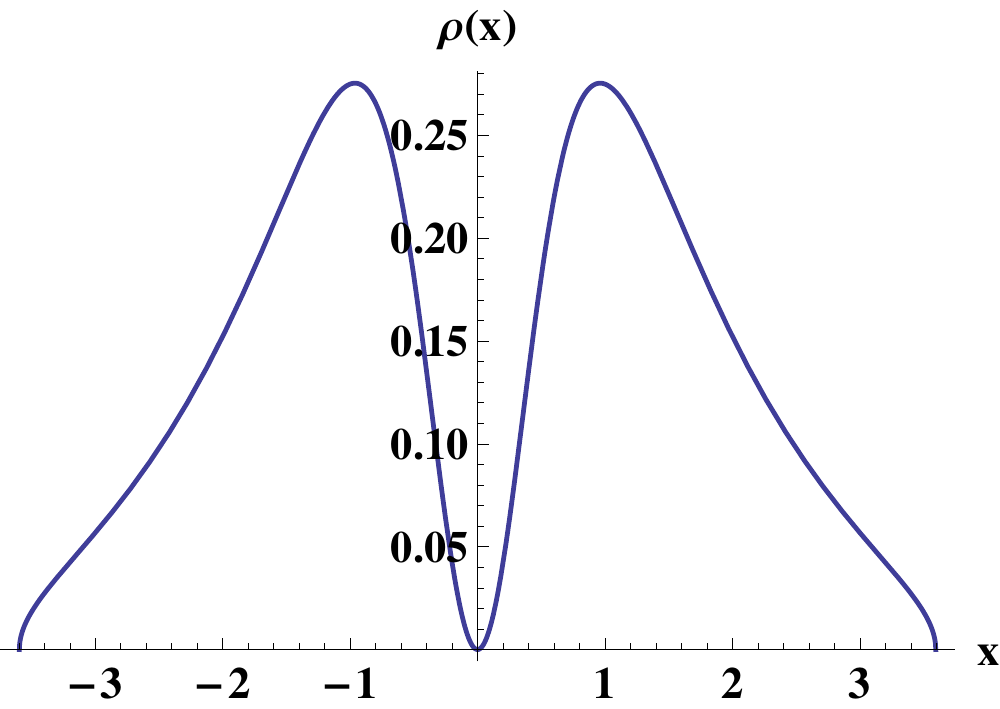}
\caption{The eigenvalue density at the critical point.}
 \label{criticalrho}
\end{figure}

Next, let us consider the condition that $\rho(x)$ is non-negative. From \eqref{rhotot} we see that $\rho$  becomes negative in an interval
when 
\be
\kappa > \kappa_c= \tau \, \frac{1+\mu^2}{1+\frac{\mu^2}{2}} \  ,
\label{crib}
\ee
where $\mu $ is determined in terms of $\tau$ and $\kappa $ by the normalization condition \eqref{norma}.
This condition indeed determines the critical value of the phase transition. Note that $\kappa_c/\tau$ is greater than one,
which implies that the transition occurs after the potential developed a double well.
We can use \eqref{crib} to find the critical $\mu $:
 \be
\mu_c^2=\frac{2(\kappa -\tau )}{2\tau-\kappa }\ .
\ee
Substituting into \eqref{norma}, we find
\be
\kappa_c=\tau +\sqrt{2\tau -1}\ .
\label{uuu}
\ee
At the critical point, the eigenvalue density becomes
\be
\rho_c(x)= \frac{(\tau -1)x^2}{\pi (1+x^2)^2} \sqrt{\mu^2-x^2}\ .
\label{rhocris}
\ee
 The  eigenvalue density is shown in fig. \ref{rhosubc}
 for $\kappa <\kappa _c$ and the critical eigenvalue density is shown in fig. \ref{criticalrho}.

The critical coupling $\kappa _c$ lies in the interval $\tau <\kappa _c<2\tau $ (in particular, this  ensures that $\mu_c^2>0$).
For example, for $\tau =2$, the potential has a double-well when $\kappa >2$. However,  eigenvalues  are still distributed along one cut until the critical value $\kappa _c\approx 3.73$, where $\mu_c \approx 3.6$. As shown below, for $\kappa >\kappa _c$, eigenvalues get distributed symmetrically in a two-cut distribution.

Another interesting limit is the case $\kappa =0$, where the density
takes the form
\be
\rho(x)= \frac{\tau -1}{\pi (1+x^2)} \sqrt{\mu^2-x^2}\ ,\qquad \mu=\frac{\sqrt{2\tau-1}}{\tau -1}\ .
\ee

\subsection{The two-cut solution}

Above the critical coupling, $\kappa >\kappa_c$, one has to search for a two-cut solution.
We assume that the eigenvalue
density has support in two disconnected regions $(-a,-b)$ and $(b,a)$, with real $a,b$ and $0<b<a$.
The density is assumed to be of the form
\be
 \rho(x) =f(x)\sqrt{(a^2-x^2)(x^2-b^2)}\ .
\label{densu}
\ee
For a two-cut ($Z_2$-symmetric) solution, the resolvent is given by
\be
\omega(z) =\frac12 \sqrt{(a^2-z^2)(z^2-b^2)} \oint_{\cal C} \frac{dx}{2\pi i}\frac{V'(x)}{z-x}\frac{1}{\sqrt{(a^2-x^2)(x^2-b^2)}}\ .
\label{ssss}
\ee
The contour ${\cal C}$ is the union of two contours surrounding the
two cuts. 
The integral can be computed by residues, and it is contributed
by the poles at $z=\pm i$.
We find
\be
 f(x) =\sqrt{x^2}\left(-\frac{ \tau }{\pi (1+a^2)^{\frac12} (1+b^2)^{\frac12} (1+x^2)} +\kappa \frac{(4+3a^2+3b^2+(a^2+b^2+2)x^2+2a^2b^2) }{2\pi (1+a^2)^{\frac32} (1+b^2)^{\frac32} (1+x^2)^2} \right).
\label{rhotot3}
\ee
The parameters $a$ and $b$ representing the endpoints of the eigenvalue distribution  can be computed by demanding two conditions:
1) normalization and 2) the asymptotic condition obeyed by
the resolvent
\be
\omega(z)\sim \frac{1}{z}\ .
\ee
One can get the equivalent condition by substituting
the above solution \eqref{densu}, \eqref{rhotot3} into the integral equation.
The integral equation then implies that  the residues at infinity coming from the two terms with coefficients $\tau $ and $\kappa $ must cancel.
This  leads to the condition 
\be
\tau (1+a^2)(1+b^2)-\frac12\, \kappa (2+a^2+b^2)=0\ .
\label{gol}
\ee
%
%
%
The normalization condition leads to the following relation between parameters
\be
1=-\frac{\tau }{2}\, \frac{ 2+a^2+b^2}{ (1+a^2)^{\frac12} (1+b^2)^{\frac12} }+\tau
 +\frac{\kappa }{4}\frac{(a^2-b^2)^2}{(1+a^2)^{\frac32} (1+b^2)^{\frac32} } \ .
\ee
Using condition (\ref{gol}), this simplifies to
\be
(1+a^2)(1+b^2)= \frac{\kappa ^2}{(\tau -1)^2} \ .
\label{masequ}
\ee
%
Equations \eqref{gol} and \eqref{masequ} can be explicitly solved for $a$ and $b$. One obtains
\be
1+a^2=
\frac{\kappa (\tau +\sqrt{2\tau-1})}{(\tau -1)^2}\ ,
\qquad
 1+b^2=
\frac{\kappa}{\tau+\sqrt{2\tau-1}}\ .
\ee
Substituting these formulas in \eqref{rhotot3}, the eigenvalue density dramatically simplifies,
\be
\rho(x)=\frac{\tau -1}{\pi (1+x^2)^2}\sqrt{x^2}  \sqrt{(a^2-x^2)(x^2-b^2)}\ .
\label{rhofinale}
\ee
This represents the exact eigenvalue distribution in the supercritical phase in the stability regime $\tau>1$.

The critical point of the transition corresponds to $b=0$. This gives
\be
\kappa _c=\tau +\sqrt{2\tau -1}\ ,
\ee
in agreement with the critical point obtained in the subcritical regime.
At the critical point,
\be
a_c^2=\mu_c^2 =\frac{2 \tau \left(\sqrt{2 \tau -1}+2\right)-2}{(\tau -1)^2}\ ,
\ee
and $\rho $ matches the critical density $\rho_c $, given in \eqref{rhocris}, obtained from the subcritical phase.


\begin{figure}[h!]
\centering
\begin{tabular}{cc}
\includegraphics[width=0.4\textwidth]{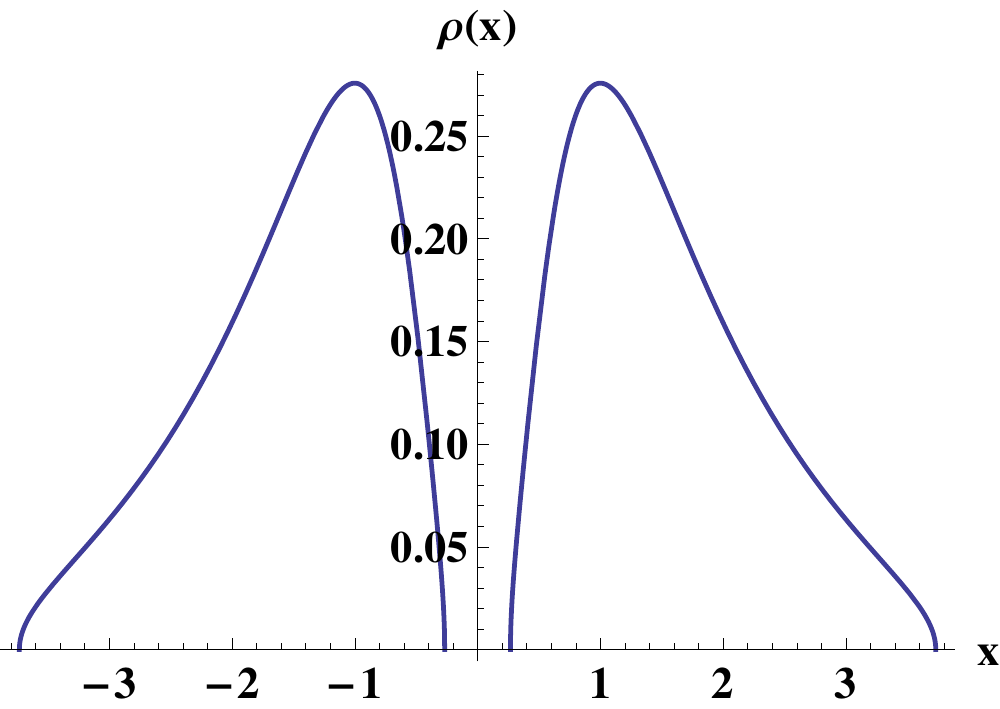}
&
\qquad \includegraphics[width=0.4\textwidth]{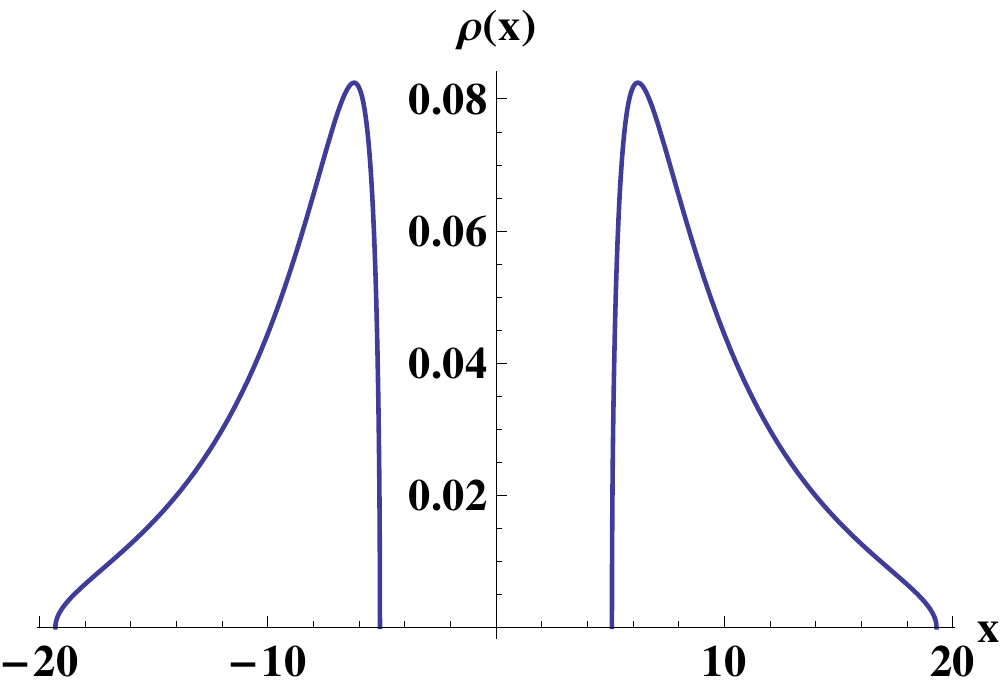}
\\
(a)&(b)
\end{tabular}
\caption
 {The eigenvalue density in the two-cut case describing the supercritical regime $\kappa >\kappa_c$. (a) $\tau=2$, $\kappa=4$. (b) $\tau=2$, $\kappa=100$, illustrating the asymptotic shape of the eigenvalue density.
}
\label{rhosuper}
\end{figure}

Finally, at large $\kappa $, the density takes the asymptotic shape
shown in fig.  \ref{rhosuper}(b), with
\be
a^2\approx \frac{\kappa }{\tau -\sqrt{2\tau-1}} .
\ ,\qquad b^2\approx \frac{\kappa }{\tau +\sqrt{2\tau-1}}\ .
\ee
Thus $a^2$, $b^2$ grow linearly with $\kappa $ with $a/b$ tending to a fixed value for given $\tau $.

\subsection{Critical behavior}

To understand the nature of the phase transition we now study the analytic properties of  the free energy $F=-\ln Z$ 
at the critical point.
The free energy can be computed in  the subcritical and supercritical regimes by using 
the one-cut and two-cut eigenvalue densities obtained above.
We shall fix $\tau $ and  increase $\kappa $ until it overcomes the critical value, where the system undergoes a phase transition.
Instead of computing the free energy, it is more convenient to consider the first derivative
\be
\frac{\partial F}{\partial \kappa } =\langle {\rm Tr} \ \frac{1}{1+M^2}\rangle =\int_{L} dx \, \rho(x) \, \frac{1}{1+x^2}\ .
\label{dBF}
\ee

\subsubsection*{Subcritical regime  $\kappa <\kappa _c$}

The integral \eqref{dBF} can be computed by residues using \eqref{rhotot} and choosing a contour surrounding the cut
from $-\mu$ to $\mu$. We obtain
\be
\frac{\partial F}{\partial \kappa }\bigg|_{\kappa <\kappa _c}=\frac{\tau \mu ^2}{2(\mu ^2+1)}-\frac{\kappa  \mu ^2 \left(\mu ^2+4\right)}{8
   \left(\mu ^2+1\right)^2}\ ,
\label{sdBF}
\ee
where $\mu =\mu(\tau ,\kappa )$ is implicitly defined  by the condition \eqref{norma} (we omit the explicit expression given 
in terms of a solution of a cubic equation).

Higher derivatives of the free energy in the subcritical phase can be computed by differentiating  \eqref{sdBF} with respect
to $\kappa $.
We need $\partial\mu^2/\partial \kappa $, which is obtained by differentiating the normalization condition \eqref{norma}. We get
\be
\frac{\partial\mu^2}{\partial\kappa }=\frac{2\mu^2(1+\mu ^2) }{2 \tau \left(\mu ^2+1\right)+\kappa  \left(\mu ^2-2\right)}\ .
\label{dmudB}
\ee
Differentiating \eqref{sdBF} and using \eqref{dmudB} we find
\bea
&&\frac{\partial^2 F}{\partial \kappa ^2}\bigg|_{\kappa<\kappa_c}=-\frac{\mu ^4}{8 \left(\mu ^2+1\right)^2}\ ,
\nonumber\\
\nonumber\\
&&\frac{\partial^3 F}{\partial \kappa^3}\bigg|_{\kappa<\kappa_c}=-\frac{\mu ^4}{2\left(\mu ^2+1\right)^2 \left(2 \tau \left(\mu
   ^2+1\right)+\kappa \left(\mu ^2-2\right)\right)}\ .
\eea

\subsubsection*{Supercritical regime $\kappa >\kappa_c$}

Let us now compute the integral \eqref{dBF}  by residues using \eqref{rhofinale}. We choose  a contour which is the union of two contours surrounding the two cuts
from $-a$ to $-b$ and from $b$ to $a$. We find the
remarkably simple formula:
\be
\frac{\partial F}{\partial \kappa }\bigg|_{\kappa >\kappa_c}=\frac{2 \tau-1}{2 \kappa }\ .
\ee
By differentiating with respect to $\kappa $, we obtain second and third derivatives
\be
\frac{\partial^2 F}{\partial \kappa^2}\bigg|_{\kappa>\kappa_c}=-\frac{2 \tau-1}{2 \kappa^2}\ ,
\qquad \frac{\partial^3 F}{\partial \kappa^3}\bigg|_{\kappa>\kappa_c}=\frac{2 \tau-1}{ \kappa^3}\ .
\ee

\begin{figure}[h!]
\centering
\includegraphics[width=0.45\textwidth]{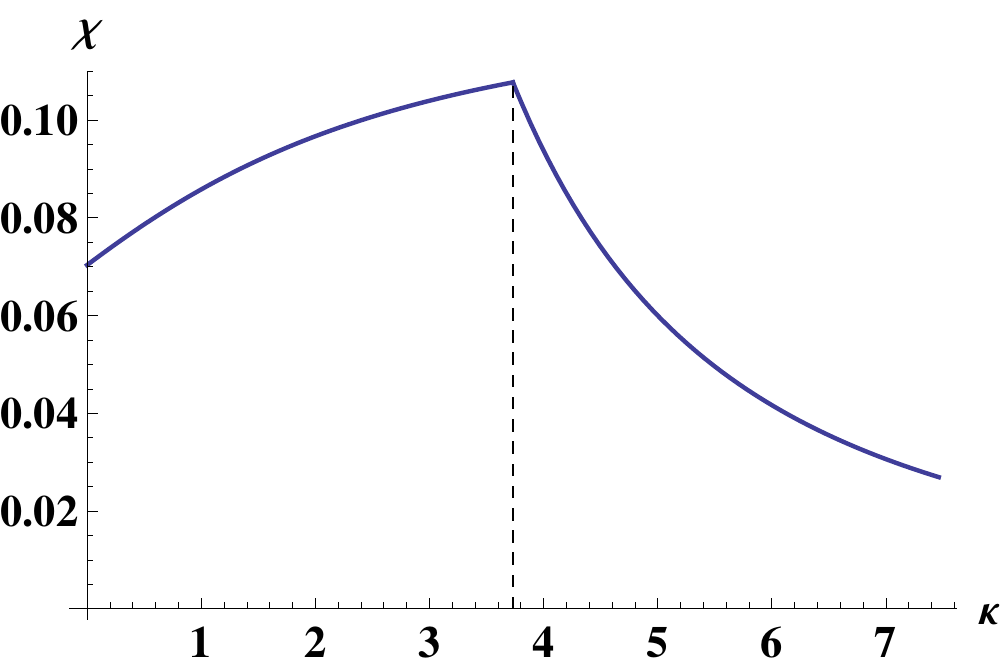}
\caption{The first derivative of the susceptibility
is discontinuous, indicating a third-order phase transition. The figure corresponds to  $\tau=2$, in which case $\kappa_c\approx 3.73$. }
 \label{sucept}
\end{figure}

We can now examine the continuity properties of derivatives of the free energy at the critical point.
For the first and second derivatives we obtain
\bea
&&\frac{\partial F}{\partial \kappa}\bigg|_{\kappa\to \kappa_c^+}=\frac{2 \tau -1}{2 \left(\tau + \sqrt{2 \tau -1} \right)}
=\frac{\partial F}{\partial \kappa}\bigg|_{\kappa\to \kappa_c^-}\ ,
\nonumber\\
\nonumber\\
&&\frac{\partial^2 F}{\partial \kappa^2}\bigg|_{\kappa\to \kappa_c^+}=
-\frac{2\tau-1 }{2\left( \tau +\sqrt{2 \tau-1}\right)^2}
=\frac{\partial^2 F}{\partial \kappa ^2}\bigg|_{\kappa\to \kappa_c^-} .
\eea
Therefore  the first and second derivatives of the free energy are continuous at the transition point.
For the third derivative, at the critical point we find
\bea
&&\frac{\partial^3 F}{\partial \kappa^3}\bigg|_{\kappa\to \kappa_c^-}=\frac{  \tau \left(2-\sqrt{2
   \tau-1}\right)-1}{2\left(\tau+\sqrt{2 \tau-1}\right)^3}\ ,
\nonumber\\
\nonumber\\
&&\frac{\partial^3 F}{\partial \kappa^3}\bigg|_{\kappa\to \kappa_c^+}=\frac{ 2 \tau-1}{\left( \tau+\sqrt{2 \tau-1}\right)^3}\ ,
\eea
and
\be
\frac{\partial^3 F}{\partial \kappa^3}\bigg|_{\kappa\to \kappa_c^+}-\frac{\partial^3 F}{\partial \kappa^3}\bigg|_{\kappa\to \kappa_c^-}=
\frac{  \tau \left(2+\sqrt{2
   \tau-1}\right)-1}{2\left(\tau+\sqrt{2 \tau-1}\right)^3}\ .
   \label{saltito}
\ee
This is different from zero for any $\tau$ in the region of stability of the theory $\tau >1$ (it has a zero at $\tau=1/2$).
Thus we conclude that the system undergoes a third-order large $N$ phase transition.
The susceptibility $\chi = - \frac{\partial^2 F}{\partial \kappa^2}$
is continuous at the transition point, but its derivative
has a jump. This is shown in fig. \ref{sucept}.

\subsection{The marginal case $\tau=1$}

The case $\tau =1$ is the marginal case for stability.
It corresponds to having $A=N+O(1)$ in the original
coupling.
The eigenvalue density can be obtained from the formulas of the previous subsections  by taking the limit $\tau\to 1$.

\begin{figure}[h!]
\centering
\begin{tabular}{cc}
\includegraphics[width=0.4\textwidth]{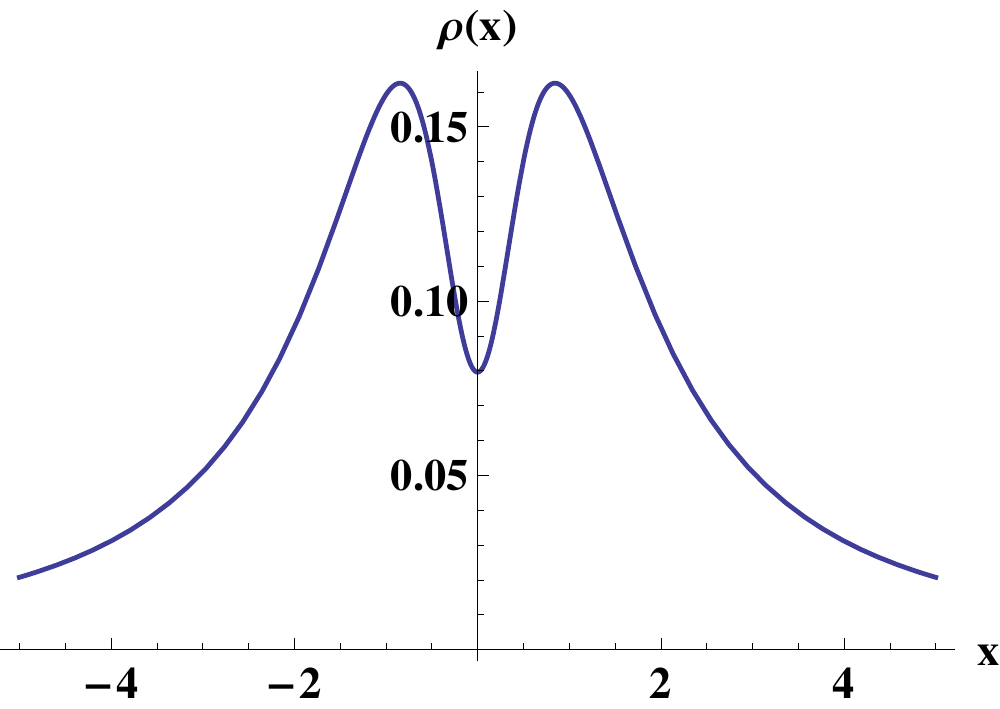}
&
\qquad \includegraphics[width=0.4\textwidth]{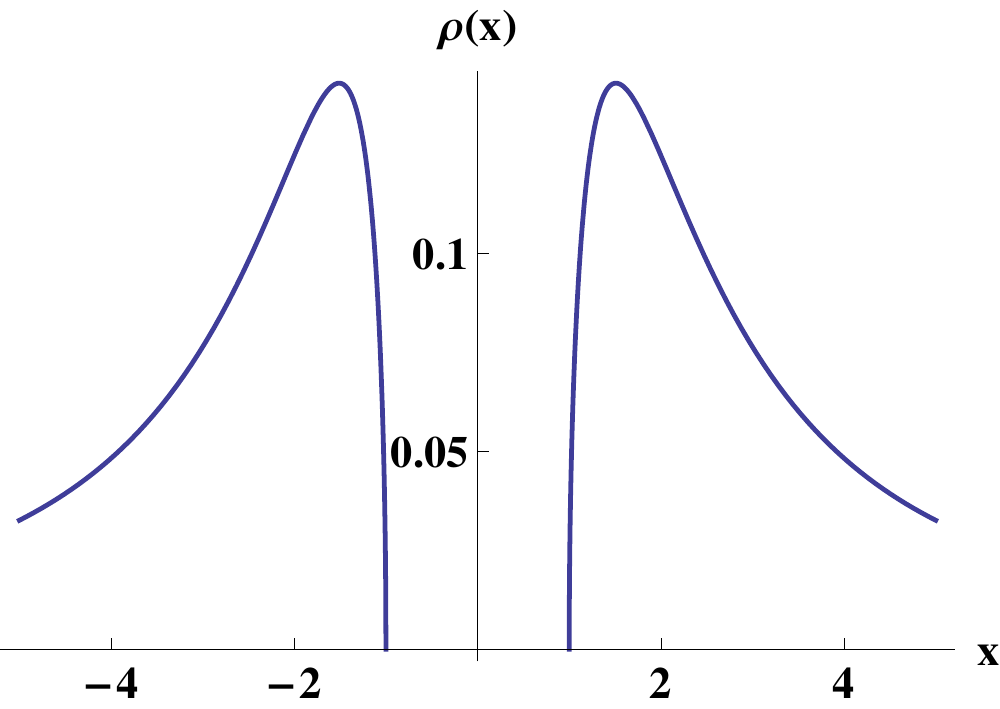}
\\
(a)&(b)
\end{tabular}
\caption
 {Eigenvalue densities in the marginal case $\tau =1$.
 The phase transition occurs at $\kappa_c=2$.
 (a) Subcritical case with $\kappa = 1.5$. (b) Supercritical case with $\kappa = 4$.
}
\label{rhomargi}
\end{figure}

In the subcritical case, $\tau\to 1 $ gives $\mu\to\infty$, that is, eigenvalues are spread from $-\infty $ to $\infty$. The critical coupling is
\be
\kappa_c = 2\ .
\ee
In the supercritical case, $a\to \infty$
and $b^2$ becomes
\be
b^2=\frac{\kappa}{2}-1 \ .
\ee
The resulting eigenvalue densities in the subcritical and supercritical case are
\be
\rho(x) =\frac{1}{\pi  \left(x^2+1\right)}+\kappa \, \frac{x^2-1}{2 \pi  \left(x^2+1\right)^2} \ ,\qquad \tau =1 \ ,\ \ \kappa \leq \kappa_c\ ,
\ee
and
\be
\rho(x) =\frac{ \sqrt{2\kappa } }{\pi  \left(x^2+1\right)^2} \, \sqrt{x^2}\  \sqrt{x^2-b^2}\ ,\qquad \tau =1 \ ,\ \ \kappa > \kappa_c\ .
\ee
One can check that they satisfy the normalization condition for any $\kappa$.
The densities are shown in figs. \ref{rhomargi}(a),(b).

At the critical point, the free energy exhibits the same non-analytic behavior as in the case $\tau>1$, with a discontinuous third derivative.
The jump is obtained from \eqref{saltito} by setting $\tau =1$.

\section{Finite $N$ partition functions}

Let us consider the 3-parameter deformed Cauchy matrix model with potential \eqref{potek} at finite $N$, {\it i.e.} before the limit $\epsilon\to 0$.
The partition function is given by
\be
Z=\frac{1}{N!}\int \frac{d^Na}{(2\pi)^N}\, \prod_{i<j}^N (a_i-a_j)^2 \, \prod_{i=1}^N 
\frac{(1+a_i^2)^\beta}{(1+a_i^2)^A (1+\epsilon + a_i^2)^\beta}\ .
\label{partis}
\ee
The integral is a complicated generalization of the Selberg's integral, with singularities at different places.
The  particular case $A=N$ was studied recently in  \cite{Santilli:2020ueh}.

In the case $\epsilon =0$, one has the Cauchy random matrix ensemble (see {\it e.g.} \cite{witte}). 
Note that for any given $A$
all  moments $\langle {\rm Tr}\ M^{2n} \rangle$ with $n\geq \frac12 + A-N$ are divergent. 
This is of course the same pathology  that is present in the Cauchy probability distribution.  This feature is  also present
in the above three-parameter deformed Cauchy matrix model.
It is worth noting that the large $N$ theory  does not suffer 
from this pathology as long as $A>N$: then all moments $\langle x^{2n}\rangle$ are finite.
The problem appears at large $N$ when $A\leq N$. 
In the marginal case $A=N$, discussed in section 2.6, already the moment $\langle x^2\rangle$ is ill-defined. 
However, we stress that there are  observables which are well defined, such as the free energy and its derivatives, computed in the previous section.

For any given $N$, the integral \eqref{partis} can be carried out explicitly
in terms of hypergeometric functions.
In the $U(1)$ case, we obtain
\bea
Z^{U(1)}&=& \frac12 \sec (\pi  (\alpha
   -\beta ))\bigg( \frac{  (\epsilon +1)^{\frac{1}{2}-\alpha } \Gamma \left(\alpha
   -\frac{1}{2}\right) }{\Gamma (\beta ) \Gamma \left(\alpha -\beta
   +\frac{1}{2}\right)}
  \, _2F_1\left(\alpha
   -\frac{1}{2},\alpha -\beta ;\alpha -\beta +\frac{1}{2};\frac{1}{\epsilon
   +1}\right)
   \nonumber\\
   &-& \frac{\pi ^{1/2} (\epsilon +1)^{-\beta }  }{\Gamma (\alpha -\beta ) \Gamma
   \left(-\alpha +\beta +\frac{3}{2}\right)} \, _2F_1\left(\frac{1}{2},\beta ;-\alpha +\beta
   +\frac{3}{2};\frac{1}{\epsilon +1}\right)\bigg)\ ,
\label{uuno}
\eea
where $\alpha\equiv A$.

It is interesting to take the limit $\beta\to\infty$, 
$\epsilon\to 0$, with fixed $B=\beta \epsilon$, where the model reduces to the theory with potential \eqref{poti}  studied here.
The partition function reduces to
\be
 Z^{U(1)}\to Z_0^{U(1)}\equiv \int_{-\infty}^\infty \frac{dx}{2\pi} \ \frac{\exp[-B/(1+x^2)]}{(1+x^2)^A}\ .
\label{azes}
\ee
This integral can be computed by taking the $\epsilon\to 0$ limit on the result \eqref{uuno}. For this, one first uses Kummer's transformations
to bring the hypergeometric functions to hypergeometric functions with argument $-\epsilon $. Then one considers the Taylor series expansion of the hypergeometric in powers of $\epsilon $ and uses the Stirling-de Moivre formula for the Gamma functions with large arguments in the coefficients of the series. Upon taking the $\epsilon\to 0$ limit with fixed $B$ and resumming the
resulting series, the hypergeometric functions become confluent hypergeometric functions and one finally obtains
\be
Z_0^{U(1)}= -\frac{\pi ^{1/2} \sec (\pi  A ) \, _1F_1\left(A
   -\frac{1}{2};A ;-B\right)}{2\Gamma \left(\frac{3}{2}-A
   \right) \Gamma (A )}\ .
\ee
One can check numerically that this is indeed the exact formula for the  integral \eqref{azes}.
Thus we have computed the partition function for our theory in the $U(1)$ case. For positive integer $A$,
$Z_0^{U(1)}$ is expressed in terms of Bessel functions.

\smallskip

In the $U(2)$ case, one obtains
\be
Z^{U(2)} =  Z^{U(1)} J\ ,
\ee
where $J$ is the integral
\be
J=\int_{-\infty}^{\infty}  \frac{da}{2\pi}\, \frac{a^2 (1+a^2)^\beta}{(1+a^2)^A (1+\epsilon + a^2)^\beta}\ .
\ee
This is also expressed in terms of $\, _2F_1$ hypergeometric functions.

\medskip 

A challenging problem is to derive a closed formula for arbitrary $N$. A standard method to compute the partition function is through orthogonal polynomials and recursion relations.
In the particular case $\epsilon =0$,  the partition function substantially simplifies and it  can be computed using Romanovski polynomials \cite{witte,Santilli:2020ueh}. 

The ensemble  \eqref{partis} with general $A,\ \beta,\ \epsilon$ and $N$ seems to have been overlooked in the literature. It would be extremely interesting  understand its different limits and phase structure.

\subsection*{Note added}

By the stereographic map of real eigenvalues to the unit circle, one constructs the  {\it unitary} matrix model which is dual to the present
Hermitian matrix model.
It was recently found \cite{Russo:2020eif} that the resulting 
unitary matrix model represents a 1-parameter deformation of
the celebrated Gross-Witten-Wadia (GWW) matrix model \cite{Gross:1980he,Wadia:2012fr} describing lattice gauge theory in
1+1 dimensions, where the coupling $B$ corresponds to $-4/g^2$, $g$ being the gauge  coupling and  the coupling $A$ corresponding to a specific characteristic polynomial insertion of the form
$\det(1+U)^A\det(1+U^\dagger)^A$. Thus the partition function of the model computes the vacuum expectation value of this gauge invariant, physical observable.
The phase transition described here also takes
place in this deformed GWW model and generalizes the GWW phase transition in the presence of an extra coupling.

\subsection*{Acknowledgments}
%
We thank M. Tierz and  K. Zarembo for valuable comments
and remarks.
We acknowledge financial support from projects 2017-SGR-929, MINECO
grant FPA2016-76005-C.

\end{document}